# Evaluating improvements on using Large Language Models (LLMs) for property extraction in the Open Research Knowledge Graph (ORKG)

Sandra Schaftner[1][0009-0008-3235-3042]

[1] sandra.schaftner@informatik.tu-chemnitz.de

**Abstract.** Current research highlights the great potential of Large Language Models (LLMs) for constructing Scholarly Knowledge Graphs (SKGs). One particularly complex step in this process is relation extraction, aimed at identifying suitable properties to describe the content of research. This study builds directly on previous research of three Open Research Knowledge Graph (ORKG) team members who assessed the readiness of LLMs such as GPT-3.5, Llama 2, and Mistral for property extraction in scientific literature. Given the moderate performance observed, the previous work concluded that fine-tuning is needed to improve these models' alignment with scientific tasks and their emulation of human expertise. Expanding on this prior experiment, this study evaluates the impact of advanced prompt engineering techniques and demonstrates that these techniques can highly significantly enhance the results. Additionally, this study extends the property extraction process to include property matching to existing ORKG properties, which are retrieved via the API. The evaluation reveals that results generated through advanced prompt engineering achieve a higher proportion of matches with ORKG properties, further emphasizing the enhanced alignment achieved. Moreover, this lays the groundwork for addressing challenges such as the inconsistency of ORKG properties, an issue highlighted in prior studies. By assigning unique URIs and using standardized terminology, this work increases the consistency of the properties, fulfilling a crucial aspect of Linked Data and FAIR principles – core commitments of ORKG. This, in turn, significantly enhances the applicability of ORKG content for subsequent tasks such as comparisons of research publications. Finally, the study concludes with recommendations for future improvements in the overall property extraction process.

**Keywords:** Large Language Models (LLMs), Scholarly Knowledge Graphs (SKGs), Knowledge Graphs (KGs), Relation Extraction, Property Extraction, Open Research Knowledge Graph (ORKG), GPT-3.5, Prompt Engineering, FAIR Principles, Linked Data, Semantic Web, Research Knowledge Representation, Information Retrieval.

## 1 Introduction

Large Language Models (LLMs) have seen rapid adoption since the release of GPT-3 in 2020 and have had a significant impact on many aspects of modern life, serving as assistants in the workplace or as creative tools for leisure activities. LLMs have also gained considerable importance in research, functioning as tools for literature searches, writing assistance, and even as subjects of investigation themselves [1]. A relatively new field of research is the integration of LLMs and Knowledge Graphs (KGs), which aims to generate useful synergies. KGs offer the advantages of accuracy and structured knowledge, while LLMs excel in contextual understanding and language comprehension [2]. These qualities make LLMs ideal tools for creating KGs based on natural language [3],[4].

However, like all tools, LLMs only work well when used correctly. For LLMs in particular, this means that only the right prompts will lead to the desired result. It is no coincidence that prompt engineering has now developed into a professional activity in its own right, because crafting good prompts brings great added value to the result.

In a fully automated process, there are also formal requirements in addition to the correctness of the content of the answers, because the answers of the LLMs must be processed directly and must therefore be parsed. This paper also addresses this challenge and shows that end-to-end automation is possible.

The remainder of the paper is organized as follows: In Section 2, a problem analysis is carried out and two research questions are posed. Section 3 outlines related work on Scholarly Knowledge Graphs (SKGs), LLMs for KG construction, and the specific task of relation extraction. Section 4 describes the materials and methods used, detailed in four subsections. Section 5 presents the results of this study, addressing and discussing the two research questions. In Section 6, suggestions for improvement for future research and for a possible implementation of ORKG property extraction are presented. Finally, Section 7 concludes with a short summary of findings.

## 2 Problem Analysis

In the field of KGs for scientific publications, the requirements are particularly high, as understanding the language of scientific texts and ensuring the precision of information extraction are critical. A well-known initiative in this area is the Open Research Knowledge Graph (ORKG) [5], which not only represents metadata but also captures the content of research publications in the form of an SKG. The ORKG is manually curated by experts and authors using a crowdsourcing approach. According to members of the ORKG team, the ORKG adheres to best practices (e.g., FAIR principles) and provides services to support the production, curation, publication, and use of FAIR scientific information [6].

However, the current manual curation approach has certain disadvantages, such as the significant time required, which hinders scalability, and the inconsistencies arising from the crowdsourcing process [7], [8]. To address these issues, Nechakhin et al. [8] conducted a study evaluating the readiness of LLMs for extracting properties, referred to as dimensions in their study. They concluded that the prompt engineering strategies few-shot prompting, and Chain of Thought (CoT) prompting did not provide better results than zero-shot prompting without applied techniques. Furthermore, when comparing LLM-generated results with ORKG gold-standard properties, they found that the "alignment with scientific tasks and mimicry of human expertise" [8] was insufficient. Consequently, they recommended using LLMs merely as suggestion tools for human experts and identified fine-tuning as necessary for achieving better results [8].

This paper aims to demonstrate that significant and highly impactful improvements in LLM results can be achieved not only through fine-tuning but also through better-formulated prompts. To this end, advanced prompt engineering techniques are applied in combination, offering a simpler and more cost-effective alternative to fine-tuning. Previous studies have shown that advanced prompt engineering can substantially improve results and eliminate the need for labor-intensive fine-tuning [10], [11]. While Nechakhin et al. [8] tested few-shot and CoT techniques in isolation, earlier studies suggest that combining multiple prompt engineering techniques yields greater improvements, including in the specific task of relation extraction [12]. Moreover, several works emphasize that the proper application of prompt engineering techniques is crucial, as small details can significantly influence outcomes [9], [10], [11]. This study consolidates insights from scientific studies and applies them to develop an optimized prompt for property extraction.

The objective of this paper is to show that advanced and properly applied prompt engineering techniques can significantly enhance results. Conversely, the findings of Nechakhin et al. [8] suggest that improperly applied techniques fail to improve LLM outcomes. To ensure optimal comparability, this study adopts the evaluation methodology of Nechakhin et al. [8], where an LLM evaluator measures the alignment and deviation of LLM results from the ORKG gold standard, and additionally, counts the number of properties generated by the LLM that match those listed in the gold standard.

Furthermore, this study demonstrates how LLM-generated properties can be matched to existing ORKG properties retrieved via the API[1]. This step effectively assigns unique URIs to the extracted properties, aligning with the FAIR principles that the ORKG is committed to [6], [13]. Assigning unique URIs helps mitigate inconsistencies, enhances resource findability, and increases the interoperability of the approach.

The **two research questions** this study aims to answer are:

*RQ1: Can the results of property extraction by LLMs be significantly improved using advanced prompt engineering techniques?*

*RQ2: Which measures can help reduce inconsistencies in ORKG properties and improve the overall property extraction process?*

Finally, it should be noted that Nechakhin et al. [8] do not explicitly clarify the relationship between LLM-extracted properties (which they call dimensions) and ORKG properties. On one hand, their study design and objectives suggest that dimensions are intended to replicate ORKG properties. On the other hand, they state that "ORKG properties are not necessarily identical to research dimensions" [8]. However, apart from metadata like authorship and publication year – which are not extracted by the LLMs due to the prompt design – their evaluation methods consistently compare dimensions directly with ORKG properties. Statements such as "This study tests the feasibility of using pretrained Large Language Models (LLMs) to automatically suggest or recommend research dimensions as candidate properties as a viable alternative solution" [8] further support this alignment.

---

[1] https://tibhannover.gitlab.io/orkg/orkg-backend/api-doc/ (19.01.2025)

# 3 Related Work

In the following three subsections, the topic of this study is situated within the current research landscape on Scholarly Knowledge Graphs (SKGs), LLMs for Knowledge Graph (KG) construction, and the specific task of relation extraction.

## 3.1 Scholarly Knowledge Graphs (SKGs)

In response to the challenges posed by the ever-growing number of scientific publications, organizing publication information in SKGs has been proposed as a solution. Unlike the widely used PDF format, SKGs are more amenable to machine processing and, through the encoding of semantic relationships via ontologies, pave the way for sophisticated analyses and advanced services, such as forecasting research trends and generating scientific hypotheses [14], [15], [16]. Furthermore, SKGs contribute to making research findable, accessible, interoperable, and reusable (FAIR) [17]. Quality criteria for SKGs can be derived from the Linked Open Data principles and the FAIR guidelines [7].

Among the SKGs widely referenced in the relevant literature is the ORKG, for example in the papers by Borrego et al. [14], Verma et al. [16], Karmakar et al. [17], and Meloni et al. [18]. According to a recent evaluation in 2024 by members of the ORKG team, the ORKG faces significant challenges, including high workload and inconsistencies [7]. Additionally, the literature identifies several further issues of existing KGs such as incompleteness [14] and lack of accuracy [5]. Recent studies suggest that leveraging LLMs in KG construction could help address these challenges, a topic explored in more detail in the next subsection.

## 3.2 LLMs for Knowledge Graph (KG) Construction

Before the advent of Large Language Models (LLMs), context posed a significant challenge for Natural Language Processing (NLP) methods [18]. Recent advancements in LLMs – recognized for their superior knowledge completion, context awareness, linguistic capabilities, and reasoning – have heightened expectations for their role in KG construction, with potential to replace traditional methodologies [1], [4], [8], [19]. Zhu et al. [3] describe LLMs as "primary tools for KG construction".

In a review paper, Pan et al. [2] outline that the combined use of KGs and LLMs can leverage synergies due to the complementary strengths and weaknesses of these technologies. While KGs excel in accuracy, decisiveness, interpretability, and knowledge structuring, LLMs demonstrate exceptional capabilities in natural language understanding, processing, generalizability, and general knowledge [2]. These attributes of LLMs not only mitigate the prevailing issue of lacking language understanding of traditional methods but also enable greater scalability and allow continuous updates for evolving KGs, unlike embedding-based methods [20].

However, studies also highlight limitations of LLMs that can affect the quality of constructed KGs. A major issue is known as hallucinations, where LLMs generate text that appears plausible but is factually incorrect [1]. Strategies to mitigate this issue include contextualizing prompts to constrain output [21] and developing restricted prompts or fine-tuning LLMs to improve controllability and interpretability of text generation [22].

Another significant challenge is non-determinism, stemming from the probabilistic nature of LLMs, which leads to non-reproducible outputs [1]. Addressing this requires designing prompts to minimize variability, as well as using in-context examples to stabilize outputs by providing consistent input patterns [23]. Additionally, setting parameters like seed and temperature appropriately can enhance deterministic outcomes [24]. Collectively, these measures improve the reliability and consistency of LLM outputs, making them more suitable for critical applications where accuracy is paramount.

While initial efforts to develop complete processes for KG construction using LLMs have been made [2], most studies focus on individual subtasks within the overall process (e.g., [3], [20], [25], [26], [27], [28]). Notably, the naming conventions and sequence of steps in LLM-based KG construction vary across studies. For instance, Pan et al. [2] divide the process into three steps: entity discovery, coreference resolution, and relation extraction. Dessi et al. [21], however, propose a more sophisticated pipeline for automatic KG construction, consisting of three stages with several substeps, though implemented with earlier NLP methods rather than LLMs. In the first stage, the extraction modules identify entities and relations between these entities from text. The resulting triples are integrated into a common representation in the second stage, known as the entity and relationship handlers. A key subtask in this phase involves mapping relations to a target ontological schema. Finally, the triple validation modules select the final set of triples. As evident, relation extraction plays a crucial role in the pipelines designed by Pan et al. [2] and Dessi et al. [21]. The next section focuses on the subtask of relation extraction, as this study positions itself within this area.

## 3.3 Relation Extraction for Knowledge Graph (KG) Construction

The task of ORKG property extraction falls under the broader scope of Relation Extraction (RE), a widely used term in the respective literature. As outlined by Kerijiwal et al. [29], relations are actually a subtype of properties, which can be categorized into object properties (also referred to as relations) and datatype properties (also known as attributes). Furthermore, the term predicates is often used synonymously with properties. In graphs, properties are represented as edges [29].

According to Jiang et al. [30], the task of RE has been significantly transformed by the advent of LLMs, as they are capable of comprehending input texts and identifying complex relationships without the constraints of predefined patterns. Various studies have already demonstrated promising results in employing LLMs for RE. Based on these findings, Wadhwa et al. [12] even conclude that LLMs should become the default approach for RE.

Jiang et al. [30] refer to RE performed by LLMs as generative RE (GRE) and distinguish three forms: open, semi-open, and closed RE. They note that most existing studies on GRE fall into the closed or semi-open categories, where the LLM is provided with a predefined list of properties from which it must choose [30]. Examples include studies by Zhu et al. [3], Wadhwa et al. [12], Yoa et al. [25], Li et al. [31], Wan et al. [32], and Wei et al. [33]. In contrast, Jiang et al. [30] define open GRE as the task of discovering a broader range of relationships with minimal predefined constraints. In addition to Jiang et al.'s [30] work, Hao et al. [34] also apply open GRE; however, they extract triples not from predefined source texts but from the inherent knowledge of the LLMs. While empirical evidence is still limited, Jiang et al. [30] advocate for open GRE over closed GRE to unlock the full potential of LLMs. They propose a workflow that first explores as many relations as possible without constraints and then refines them in a subsequent step.

The property extraction task described by Nechakhin et al. [8] can also be classified as open GRE since no predefined list of properties is provided. However, Nechakhin et al.'s study [8] lacks the refinement step suggested by Jiang et al. [30]. In the present study, the refinement will be applied in the property matching phase in which the extracted properties will be aligned with existing ORKG properties. This phase serves as a filtering mechanism ensuring applicability and quality, resembling the refinement step outlined by Jiang et al. [30].

# 4 Materials and Methods

The following subsections describe the underlying materials (Subsection 4.1) as well as the methods used in this study (Subsections 4.2, 4.3, and 4.4). All materials and scripts employed for method implementation and evaluation are publicly accessible on the project's GitHub page[2].

## 4.1 Material: Preliminary comparison study and evaluation dataset by Nechakhin et al.

This study builds on the work titled *"Evaluating Large Language Models for Structured Science Summarization in the Open Research Knowledge Graph"* by Nechakhin et al. [8]. It replicates the original study while in parallel implementing and validating proposed improvements. Consequently, much of the initial material originates from Nechakhin et al.'s study [8], hereafter referred to as the preliminary study.

The primary resource for replicating the study was the dataset published by Nechakhin et al.[3] This dataset, provided as a CSV file, includes information on 153 ORKG contributions that serve as the subject of investigation, as well as the experimental results from Nechakhin et al.'s study [8]. The columns of the dataset are described by the authors on the dataset's download page[3].

---

[2] https://github.com/SandraSchaftner/orkg_property_extraction_using_gpt-3.5 (20.01.2025)

[3] Available at https://data.uni-hannover.de/dataset/orkg-properties-and-llm-generated-research-dimensions-evaluation-dataset (19.01.2025)

For the present study, as in the preliminary study, the data in the *research_problem* column was used as the subject of investigation, and the data in the *orkg_properties* column was treated as the gold standard. Additionally, the columns *gpt_dimensions*, *mappings*, *alignments*, and *deviations* were used for comparisons with the preliminary study.

Since Nechakhin et al. [8] only published results for GPT-3.5-Turbo – which outperformed the other tested LLMs, Llama 2 and Mistral – this experiment also employs GPT-3.5-Turbo accessed via the OpenAI API to allow direct comparison and extension of the results. As the specific model version was not indicated in Nechakhin et al.'s paper, this study tested two possible versions of GPT-3.5-Turbo: "gpt-3.5-turbo-0125" (released in January 2024) and "gpt-3.5-turbo-1106" (released in November 2023).

Departing slightly from the preliminary study, three rows from the dataset were extracted for use as few-shot examples in an optimized prompt. Consequently, the gold standard and comparison dataset for this experiment contain only 150 contributions. To avoid bias, the extracted rows were randomly selected by taking one row at the end of each third of the table, corresponding to the contributions with IDs 50, 101, and 152.

The system prompts from the preliminary study were used without making any changes for property extraction as well as for evaluation in terms of alignment, deviation, and the number of mappings. These prompts served both to replicate the preliminary study's results and as a foundation for experiments with modified versions of the prompts.

## 4.2 Method: Advanced prompt engineering techniques applied

In the preliminary study, the three prompting strategies zero-shot, few-shot [35] and Chain of Thought (CoT) [36] were tested. According to Nechakhin et al., the results demonstrated that "the analysis shows that the utilization of more advanced prompting techniques did not necessarily result in superior outcomes, which leads us to believe that our original zero-shot prompt is sufficient for our task's completion" [8]. However, the paper does not provide detailed results comparing the prompting strategies. The conclusion that prompt engineering did not improve the results, which was reproduced in this study using the same prompts, contradicts the broader consensus in research that well-applied prompt engineering can significantly enhance the performance of LLMs.

The author of this paper hypothesizes that the prompt engineering techniques in the preliminary study were not applied correctly. Possible reasons for the poor performance of these strategies, informed by relevant scientific insights, are discussed below. Recommendations for improvements derived from the literature are then applied to develop an optimized prompt for the specific task of property extraction.

One of the most well-known prompting techniques is few-shot learning, also referred to as in-context learning (ICL). This involves including examples in the prompt to condition the model's responses and improve its understanding of the task. [1], [37], [38]. Numerous studies have demonstrated the benefits of few-shot prompting. For example, Brown et al. [35] showed already in 2020 that few-shot prompting consistently outperformed zero-shot prompting across all experiments with GPT-3. Specifically, for the task of relation extraction, two studies also demonstrated the substantial advantages of using few-shot learning [12], [32]. However, as other studies have highlighted, selecting the right examples and paying attention to details is crucial [9], [10], [11], as even small differences can lead to significant changes in outcomes [10], [23]. Zhao et al. [10] investigated possible causes for the large impact of example selection and order, identifying one key issue as Majority Label Bias, defined as a bias "towards answers that are frequent in the prompt" [10]. An analysis of the *orkg_properties* column in the dataset reveals that every one of the 153 ORKG property lists contains the property *research problem*. This is likely because, when creating a new contribution, the property *research problem* was pre-filled prominently in the form until a redesign of the webpage, as can be seen in an earlier demonstration video[4]. Additionally, the ORKG project's modeling best practices state that each contribution should always be assigned a research problem[5]. Consequently, the three examples selected for the optimized few-shot prompt in this study, taken from the dataset, all contain the property *research problem*. This led to nearly all properties extracted by the LLM also including *research problem*, as shown in this study's results. So, by incorporating examples that reflect the Majority Label Bias, the optimized prompt exploits this tendency to achieve more aligned outputs.

Furthermore, the three examples of the optimized prompt also include some relatively frequent properties in the gold standard, such as *Analysis* and *Data used*, which appear 12 and 11 times, respectively. In contrast, the examples used in the few-shot prompt in Nechakhin et al.'s study [8] did not include *research problem*. This highlights the poor selection of examples for their experiment.

Another well-known prompting strategy is CoT [36], [39], [40], which encourages the model to reason step by step. Kojima et al. [40] demonstrated that even a simple instruction like "Let's think step by step" before each answer significantly improves response quality. Wadhwa et al. [12] showed that for relation extraction, combining few-shot learning with CoT yields better results than using few-shot learning alone. Based on these insights, this study unifies the few-shot and CoT prompts from the preliminary study into a combined prompting strategy.

In addition to these strategies, several other recommended techniques for improving LLM responses were applied in the optimized prompt (c.f. Figure 4.2). These include adding a persona description as a prefix and using delimiter characters (e.g., ###) to separate examples from instructions [1] [42]. Furthermore, OpenAI advises providing detailed instructions and, if possible, specifying the desired output length [42]. This was implemented in the optimized prompt with the following sentences: "*The dimensions should contain more general properties like 'research problem,' 'method,' and 'study area,' as well as more specific properties. Your final result*

---

[4] https://youtu.be/8rqkQe_2lUE?si=QdhIPdU7AzEcT6k2 (19.01.2025)
[5] https://orkg.org/help-center/article/42/Modeling_best_practices_for_resources_and_properties (19.01.2025)

*should contain between 4 and 8 properties.*" The benefits of specific instructions were also demonstrated in the study by Trajanoska et al. [42]. However, it should be noted that for real-world applications, Jiang et al. [30] recommend specifying a larger number of required properties and refining them subsequently, a process that is implemented in this approach through property matching, which will be introduced in the next subsection. For the part of the property extraction, the specified range of 4 to 8 properties was chosen to make the results directly comparable with the gold standard, which has an average of 5.7 properties per contribution, with a minimum of 2 and a maximum of 18.

> Hello, you are my (1) very intelligent and helpful assistant today. You will be provided with a research problem and your task is to list research dimensions that are relevant to find similar papers for the research problem. (2) Think step by step. The dimensions (3) should contain more general properties like "research problem", "method" and "study area" as well as more specific properties. Your final result (3) should contain between 4 and 8 properties.
> Respond only in the format of a python list. (4) Delimited by ### you can find (5) three successfully completed task examples.
> ###
> Research problem: "Mapping of mineral-bound buddingtonite (dominant ammonium) in hydrothermally altered rocks from Nevada" Research dimensions: ["Data used", "research problem", "Analysis", "reference"]
> Research problem: "Sustainable supply chain" Research dimensions: ["research problem", "Resolution methods", "Uncertainty", "Sustainability factor", "Objectif function"]
> Research problem: "Persistent Identification" Research dimensions: ["research problem", "Entity type", "provides API", "uses identifier system", "Metadata schema"]
> ###

Figure 4.2. Optimized prompt for the present use case of extracting properties based on a research problem. The relevant improvements are highlighted in gray and numbered: (1) Persona description, (2) CoT strategy, (3) clear and detailed instructions, (4) use of delimiters, (5) relevant few-shot examples.

Finally, it is important to note that prompts are rarely perfect on the first attempt and typically require iterative refinement [43]. For instance, Kommineni et al. [23] describe in their paper that they continually improved their initial prompt until achieving an optimal version. This iterative approach was also adopted in this study.

## 4.3 Method: Property matching to ORKG property URIs

Matching properties to unique URIs is the next step in the KG construction process following property extraction, often referred to as "relation mapping" [21] [44]. To avoid confusion with the term "mappings" used by Nechakhin et al. [8] for comparing the number of similar properties to the gold standard, this study adopts the term "matching" for aligning properties to URIs, which is also used in the literature [44]. Through matching, the extracted properties – initially just words – are linked to unique URIs, giving them semantic meaning within the chosen ontology while contributing to consistency and unambiguity [18], [21], [44].

Property matching is also part of the manual approach employed by the ORKG. When creating a new contribution, users are encouraged to reuse properties already present in the ORKG rather than creating new ones[4,5]. However, many contributors appear not to follow this guideline, as browsing the ORKG properties reveals numerous duplicates. For example, searching for the property *method* returns more than 20 IDs representing the same concept[6], without considering plural forms like *methods*, compound terms like *used method* or *test method*, or related words such as *methodology*. This inconsistency, mentioned by D'Souza et al. [7] and Nechakhin et al. [8], is clearly evident in this example and complicates structured, machine-actionable comparisons, one of the key features of the ORKG [8], [13].

The matching method used in this study is relatively straightforward and has potential for further refinement. As shown in Figure 4.3, the Python script generates multiple slightly modified variants of the extracted properties. Each variant is searched via a literal match in a collection of over 10,000 existing ORKG properties fetched from the ORKG REST API[1]. The first match found is returned.

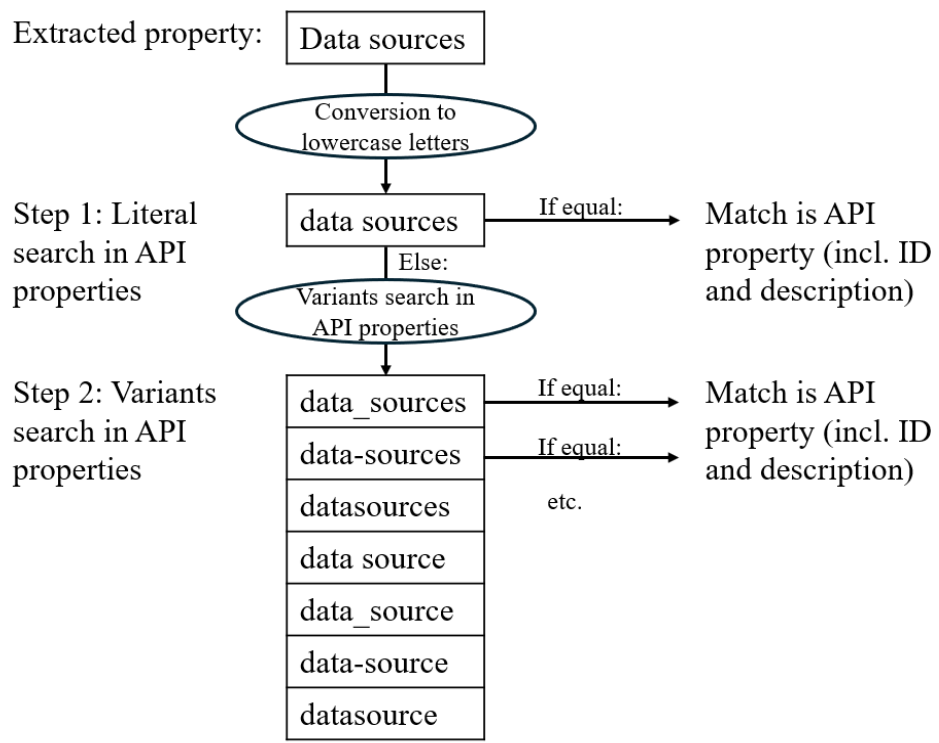

Figure 4.3. Example of script-based generation of multiple variants of a property for matching to existing ORKG properties fetched from the ORKG REST API[1]. If, for example, the LLM identifies *Data sources* as a property, the illustrated word variants are generated, each of which is then subjected to a literal search within the existing ORKG properties.

---

[6] https://orkg.org/search?q=method&types=predicate (19.01.2025)

To fully leverage the resulting advantages, such as consistency and unambiguity, the current list of over 10,000 properties in the ORKG would need to be consolidated in a real-world application. For cases like the aforementioned property *method*, where multiple semantically identical properties exist, the consolidation process should select the property that best aligns with the ORKG best practices[5], such as adhering to singular, short, and general labels and providing a clear, general description. All other properties should then be replaced by this optimal property.

## 4.4 Method: Evaluation

The evaluation of LLM results is conducted in exactly the same manner as in Nechakhin et al.'s preliminary study [8]. The primary reason for this approach is to ensure maximum comparability. Additionally, the author considers the evaluation method used by Nechakhin et al. [8] to be effective, as it does not rely on a strict (exact matching) evaluation against the gold standard but instead employs a softer matching evaluation with alignment and deviation. This approach has been recommended by other studies as well [12], [30]. Specifically, Nechakhin et al. [8] instruct the LLM GPT-3.5-Turbo which is used as a judge, to rate the semantic alignment or deviation between two lists – the LLM extracted properties and the gold standard properties – on a scale from 1 to 5.

The combination of alignment and deviation provides an inherent control mechanism for the overall accuracy of the evaluation.

Furthermore, no need was identified to modify the evaluation method, as it is assumed that the surprising result of Nechakhin et al. – that the prompting techniques few-shot and CoT did not show improvements – likely stems from other causes already discussed in Subsection 4.2. The evaluation method itself appears to be well-designed, and the use of LLMs for evaluating LLM responses is supported by several other studies. Nechakhin et al. cite examples of studies that use LLMs for evaluating translation and summary quality [8]. Similarly, in the domain of constructing KGs, there are studies that utilize LLMs for evaluation [23], including specifically for the task of relation extraction [30].

To measure the significance of differences between the various prompting strategies, the sign test is applied. The hypotheses tested include: (1) that the few-shot and CoT strategies each yield better results than zero-shot; (2) that the combined prompting strategy delivers better results than zero-shot, few-shot, and CoT; and (3) that the optimized prompts outperform all other prompting strategies. The sign test is applied to the evaluation results regarding alignment, deviation, and the number of mappings.

## 5 Results and Discussion

In the following two subsections, the research questions RQ1 and RQ2, presented in the introduction, are addressed based on the empirical results and discussed in the context of relevant literature.

## 5.1 RQ1: Evaluation of different prompting strategies

The evaluation of different prompting strategies replicated the methodology of the preliminary study by Nechakhin et al. [8] and extended it to additional prompting strategies. Before delving into a detailed comparison of these strategies, it should be noted that the findings of Nechakhin et al.'s preliminary study [8] - for no clear reason – could only be partially replicated. For the extracted GPT dimensions and the evaluations of deviation and the number of mappings, comparable results to the CSV file from Nechakhin et al.[3] were achieved. However, the alignment evaluation scores could not be replicated, even after multiple attempts using the exact same zero-shot prompts described in the preliminary study.

The alignment scores listed in the CSV file from the preliminary study are on average approximately one scale point higher than the replicated results. For example, the alignment score for ID 3 (paper title: *Km4City ontology building vs data harvesting and cleaning for smart-city services*, research problem: *Smart city ontology*) is listed as 4, indicating "substantial semantic alignment". However, an examination of the gold-standard ORKG properties and GPT dimensions shows no apparent basis for this score:

- ORKG properties: ['research problem', 'Linked Ontology', 'Ontology domains']
- GPT dimensions: ['Urban planning', 'Internet of Things', 'Data analytics', 'Sustainable development', 'Wireless communication', 'Energy management', 'Citizen participation', 'Traffic management', 'Environmental monitoring', 'Infrastructure management']

Although occasional errors by LLMs are well-documented, the discrepancies observed in this instance are not isolated cases. At the same time, the LLM generally demonstrated a tendency to assign scores correctly in the evaluations of other strategies.

The cause of the significant alignment score deviations in Nechakhin et al.'s preliminary study remains unclear. A plausible explanation, though unlikely based on the current experiment, is the use of a different version of the GPT-3.5-turbo model. The preliminary study does not specify which version was used, so this study tested both "gpt-3.5-turbo-0125" (January 2024) and "gpt-3.5-turbo-1106" (November 2023). While there were minor differences between the versions, these variations do not account for the large discrepancies in alignment scores.

However, comparing the two versions revealed the interesting result that the older model, "gpt-3.5-turbo-1106", made no errors in property extraction, even across multiple runs, while the newer model more frequently produced errors in the output format and failed to adhere to the required Python list format, complicating the parsing of the outputs. Conversely, in the evaluation tasks, the older model produced more output errors. These findings align with prior studies, which report task-specific performance variations across different GPT-3.5-turbo releases. For example, Frey et al. [45] found that the November 2023 release performed best for one KG construction task but worst for another when compared to March and June 2023 releases. Similarly, Chen et al. [46] observed substantial task-specific performance differences across GPT-3.5-turbo releases in seven diverse tasks.

For this study, the extraction errors produced by "gpt-3.5-turbo-0125" were deemed more critical than the evaluation errors of "gpt-3.5-turbo-1106". The extraction errors from "gpt-3.5-turbo-0125" affected two contributions, invalidating all evaluations for these papers, whereas with "gpt-3.5-turbo-1106" there was only one single error in the test run when evaluating the number of mappings. Consequently, results from "gpt-3.5-turbo-1106" were used for further analysis.

The findings confirm Nechakhin et al.'s observation that the applied few-shot and CoT prompts did not produce significantly better results than the applied zero-shot prompts. However, this "absence of discernible performance improvement" [8], as Nechakhin et al. describe it, does not imply that prompt engineering has no utility for this application. As discussed in Section 4.2, it is hypothesized that the prompt engineering techniques in the preliminary study were not applied optimally. This hypothesis is supported by the results presented partly in Figure 5.1 and completely in the Appendix.

As the heatmaps show, neither the few-shot nor CoT prompts yielded significantly better results than the zero-shot prompts, except in the CoT evaluation for deviation. Combining the two techniques into a single combined prompt resulted in very to highly significant improvements across all three evaluations compared to zero-shot prompts. The combined prompt also significantly outperformed the individual few-shot and CoT strategies in two of three evaluations. The optimized prompt achieved the best results, yielding highly significant improvements across all evaluations compared to all other strategies.

Based on these results, the research question, "*Can the results of property extraction by LLMs be significantly improved using advanced prompt engineering techniques?*" can be confirmed. The optimized prompt demonstrated that the techniques discussed in Section 4.2 led to significantly better outcomes in property extraction. Additionally, the optimized prompt achieved an average of 5.0 extracted properties per contribution, close to the gold standard average of 5.7. In contrast, the other strategies produced averages ranging from 7.1 to 8.9.

Nevertheless, it is important to note that while the optimized prompt achieved significantly better results, the average scores still leave room for improvement. This observation aligns with Wadhwa et al.'s [12] assertion that open-ended relation extraction without predefined options complicates evaluation. Flexible natural language responses may still be accurate and correct even if they deviate from the target [12]. This issue is evident in the evaluation of the number of mappings, where even the optimized prompt achieved only 1.2 mappings per contribution on average, representing 24% of the average 5.0 extracted properties per contribution. These findings are consistent with Wadhwa et al.'s [12] expectations that such evaluations can lead to inaccurate and overly pessimistic results.

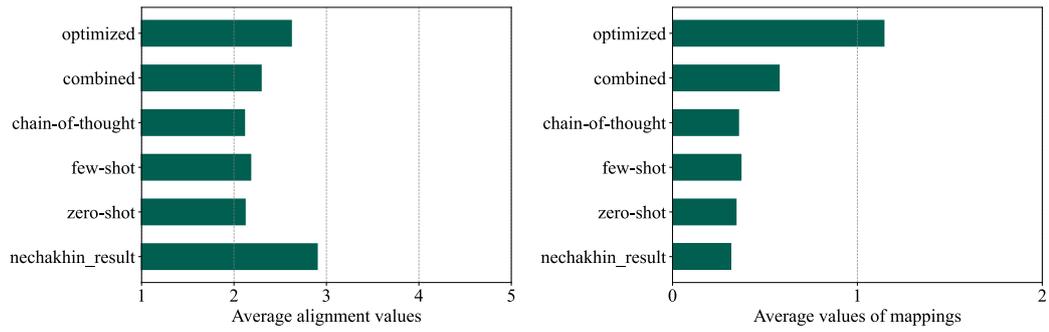

Figure 5.1. a)
Bar charts of the evaluation scores for alignment and number of mappings. Regarding the alignment bar chart, it should be noted that the results of the preliminary study, referred to as *nechakhin_result*, could not be reproduced even approximately when using the same prompts (see the discussion in the first paragraphs of Section 5.1).

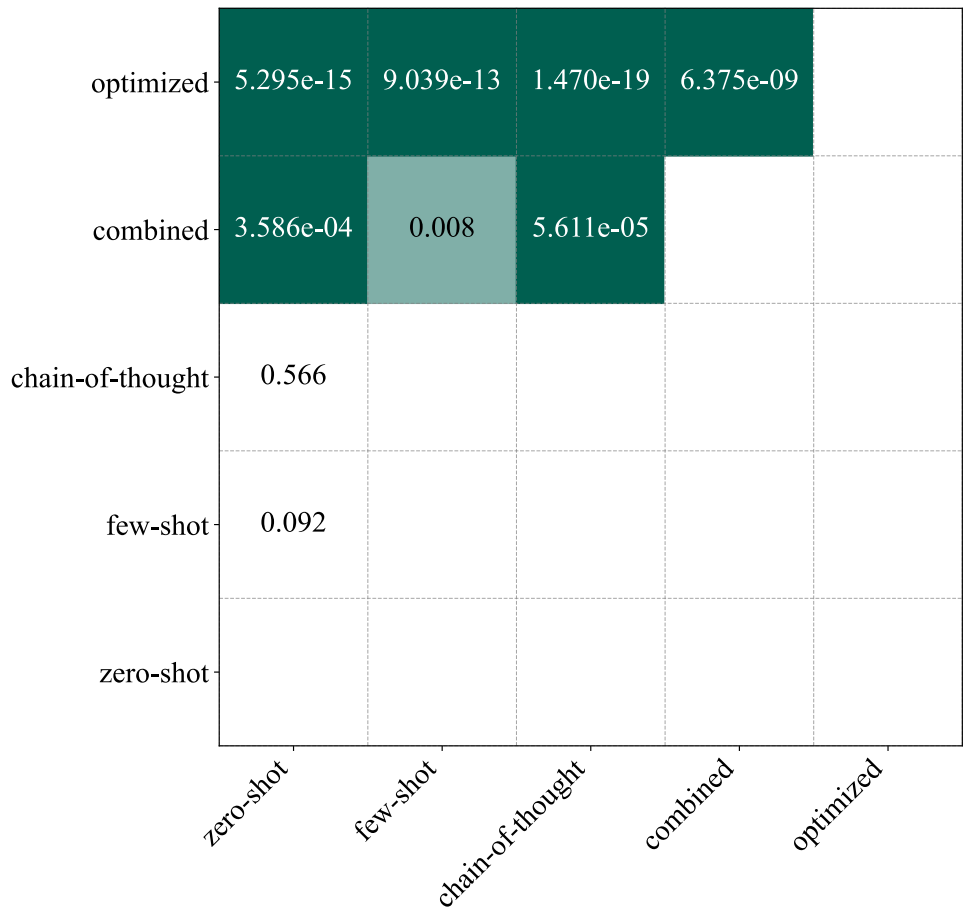

Figure 5.1. b)
Heatmap displaying the p-values from the sign tests for the results on alignment. The heatmap is to be interpreted as follows: *[The strategy listed on the vertical axis on the left]* is better by the p-value *[p-value]* than *[the strategy listed on the horizontal axis below]*. Only the cells that were calculated are filled in.
Complete set of three bar charts and three heatmaps for the evaluation regarding alignment, deviation, and number of mappings, see Appendix.

## 5.2 RQ2: Ontology matching to address inconsistencies and further improvements

With RQ2, the question was raised which measures could help minimize the existing inconsistencies in ORKG properties and improve the overall property extraction process. The response is structured around three aspects: 1) minimizing inconsistencies, 2) providing expanded context as input for LLMs, and 3) redesigning the process with predefined properties. While improvement 1) is implemented in the present study, aspects 2) and 3) are discussed here only conceptually, as these overarching improvements impact the entire process.

*Minimizing inconsistencies*

To minimize inconsistencies, this study proposes property matching as an additional step alongside the property extraction conducted in the preliminary study. Within the property matching process, the extracted properties are aligned with the appropriate URIs of existing ORKG properties. As discussed in Section 4.3, however, the effectiveness of this matching depends significantly on the quality of the existing ORKG property base, which must be free of duplicates and inaccuracies. An effective elimination of inconsistencies further necessitates the proposed consolidation of the current ORKG properties.

The property matching process itself yielded promising results in this study. Although the matching implementation was relatively straightforward – comprising direct comparisons of phrases with substitutions for characters like underscores, hyphens, and spaces, as well as conversion to singular forms – the optimized prompt still achieved an average match rate of 40%. There is considerable potential for a more sophisticated implementation that could further increase this proportion.

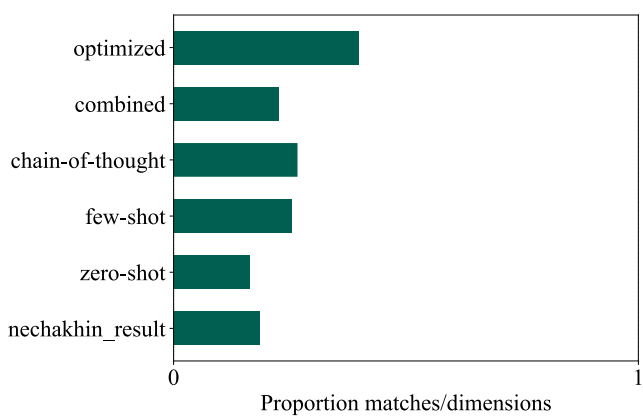

As shown in Figure 5.2, the match proportion for other prompting strategies, as well as for the dimensions extracted in Nechakhin et al.'s study (read from the provided CSV file[3]), was significantly lower. Based on these findings, it can be concluded that the optimized prompt also outperforms the other prompting strategies in this respect.

Figure 5.2. Bar chart showing the proportion of matches to existing ORKG properties relative to the total number of extracted properties per strategy.

In particular, the aspect of similarity to ORKG properties is evident in this result, as it demonstrates that a substantial portion of the properties extracted using the optimized prompt are already established as ORKG properties. Examples of properties and matches for individual contributions using the optimized prompt are presented in Table 5.2.

In contrast, the zero-shot prompting strategy and the results from Nechakhin et al. [8] – which also used the zero-shot prompting strategy – performed the worst in comparison (cf. Figure 5.2). This finding further supports the conclusion that the deviation of this strategy from the gold standard is the highest.

| Extracted properties | Matches to ORKG properties |
| --- | --- |
| **Example 1:** | |
| *research problem* | {"id": "P32", "label": "research problem", "description": "Extracting app features and corresponding sentiments for requirements engineering"} |
| *Data sources* | {"id": "P123036", "label": "data sources", "description": null} |
| *Algorithm* | {"id": "P2001", "label": "Algorithm", "description": "A finite sequence of rigorous well-defined instructions, typically used to solve a class of specific problems or to perform a computation."} |
| *Evaluation metrics* | {"id": "P41532", "label": "Evaluation metrics", "description": "Relates the contribution to the used evaluation metrics to assess the contribution. (e.g. Precision, Recall, Hits@N, AUC etc.)"} |
| *User preferences* | |
| **Example 2:** | |
| *research problem* | {"id": "P32", "label": "research problem", "description": "Extracting app features and corresponding sentiments for requirements engineering"} |
| *Data integration* | |
| *Simulation model* | |
| *Real-time monitoring* | |
| *Predictive maintenance* | {"id": "P68025", "label": "Predictive maintenance", "description": null} |
| **Example 3:** | |
| *research problem* | {"id": "P32", "label": "research problem", "description": "Extracting app features and corresponding sentiments for requirements engineering"} |
| *Technology* | {"id": "P43088", "label": "Technology", "description": null} |
| *Applications* | {"id": "P15154", "label": "Applications", "description": null} |
| *Experimental methods* | |
| *Results* | {"id": "P6001", "label": "results", "description": "Description of the main findings of the study, based on the outcome measures investigated. When multiple outcome measures are involved, each result is numbered and assigned to a different subsection. When statistical significance values were reported, they are inlcuded. "} |

Table 5.2. Three examples of extracted properties and their matches to existing ORKG properties, based on three contributions analyzed by the optimized prompt. Note: No entry in the right column means no match found by the algorithm.

*Expanded context as input for LLMs*

A key improvement involves expanding the context provided to the LLM as input for its task. Several studies have suggested and evaluated this approach as beneficial [23], [47], [48]. For the present use case, the author suggests that including the abstract and methods section of the studies as context would be particularly effective. Using only the research problem often yields unsatisfactory results because it is sometimes just a single word or phrase. Even human experts would likely struggle to produce a precise list of properties with only terms like supply chain, aiming, orientation, or nanothermometer (examples from this and the preliminary study) as context. Notably, the human experts whose results were used as the gold standard in Nechakhin et al. [8] had access to the full research papers when generating the benchmark properties.

An expanded context is also essential if subsequent steps in KG construction are to be automated. For this use case, such a step might involve linking the extracted properties to object entities. The tables used in ORKG comparisons illustrate this process well: properties form the row labels, contributions serve as the column labels (subject entities), and object entities populate the inner cells. For example, for the property *Study Area*, the corresponding cells might contain entries like "Europe", "North America", or "Hanover". However, determining such object entities often requires more context than the research problem alone. For certain properties, such as *data source* or *data coverage*, even the abstract might be insufficient, making the inclusion of the methods section as additional context advisable.

*Redesigned process with predefined properties*

The results of matching LLM-extracted properties to existing ORKG properties demonstrate that while feasible, this process is not trivial. The author questions whether this approach is optimal and proposes a conceptual two-step process instead.

In the first step, a predefined set of properties – such as *research field, research problem, method, evaluation data set* and *evaluation benchmark* – , similar to the templates used in the ORKG, would be provided and evaluated for applicability to a specific contribution. Properties deemed irrelevant for that specific contribution by the LLM would be excluded. In the second step, the LLM would be allowed to freely suggest additional relevant properties, as was done in the current study.

This approach offers several advantages. First, the predefined properties would promote greater consistency across all contributions, which is particularly useful for comparisons, as it ensures a larger number of common properties for direct comparison. Furthermore, these predefined properties would already be linked to appropriate URIs, eliminating the challenging and time-consuming step of URI matching for these properties. Simplifying the process with predefined properties and their associated URIs would also likely reduce errors compared to the method proposed by Nechakhin et al. [8].

In conclusion, the proposed two-step process – combining predefined properties with the flexibility to extract additional properties – could significantly improve the consistency, efficiency, and reliability of property extraction for KG construction.

## 6 Future Work

In terms of further research and possible implementation of LLM-based property extraction for the ORKG, there is still potential for several improvements. An essential and fundamental requirement for effective property matching is the consolidation of the ORKG ontology to avoid duplicate representations of concepts through multiple URIs. In light of recent efforts towards overarching standards and recommendations, such as those promoted by the German National Research Data Infrastructure (NFDI) initiative [17], the ORKG ontology could be adapted to align with NFDI guidelines as part of this consolidation effort.

For real-world applications of property extraction and mapping, integrating a mechanism to allow LLMs to propose new properties for the ontology could further enhance the system. These newly proposed properties would require approval from human experts to ensure consistency and quality. This approval process, however, is expected to be far less labor-intensive than the current manual extraction of all properties from research publications.

Further possibilities for improvement in future work include incorporating more context into the input [23], [47], [48], and generating a larger number of properties, which can subsequently be filtered [30]. Future studies could also explore the use of alternative LLMs. For instance, in a study focused specifically on relation extraction, OpenChat-3.5 (7B) outperformed GPT-3.5-Turbo and other large models like LLaMA-2-70B across most datasets [30].

When designing a comprehensive KG construction pipeline, it is crucial to address inherent LLM issues such as hallucinations and non-determinism. Jiang et al. [30] proposed a framework for factualness checks to mitigate these challenges. In the current use case of open-ended relation extraction without entity extraction, distinguishing between creativity and hallucination remains challenging. However, with the inclusion of entity extraction, the full triples could be more effectively evaluated for factualness.

A feasible improvement in this study's context is validating and correcting the output format. For example, with version "gpt-3.5-turbo-0125", there were instances where the prompting strategy combined failed to adhere to the required output format of a python list. In such cases, even script-based parsing of the LLM response could not produce usable results. A potential solution is employing a second LLM to convert the output from the first LLM call into the correct format. Preliminary experiments confirmed that this approach works, such as when the first LLM outputs a numbered list without brackets, which the second LLM can properly format as a python list.

This tactic, referred to as output guardrails, is widely applied also outside of scientific programming and is recommended by OpenAI [49]. It involves using a secondary LLM or fine-tuned pre-trained language model to check various aspects of the output, such as hallucinations, factual correctness, and syntax. Syntax checks, in particular, should be applied to any LLM response prone to formatting errors [49].

The development of a complete pipeline for extracting full triples, incorporating these improvements, is currently underway by the author of this paper.

# 7 Conclusions

This study demonstrates that prompt engineering techniques, when applied correctly, can significantly improve the performance of LLMs in the task of property extraction. Additionally, implementing a matching step to align extracted properties with existing ORKG properties enhances the process by addressing issues of inconsistency and lack of uniformity in properties. This matching step also improves the quality of results by ensuring that the extracted properties align with a predefined ontology, allowing for criteria such as generalizability to be systematically enforced.

# Acknowledgment

The author thanks Dr. Sheeba Samuel and Jan Ingo Haas from the Chemnitz University of Technology for their valuable comments and feedback on this paper.

Moreover, the author gratefully acknowledges Nechakhin, D'Souza, and Eger for publishing the dataset[3], prompts, and experimental setup. This made it possible to replicate, understand, and build upon their study. I hope that through my ideas and findings presented in this paper, as well as my publicly available program code, prompts, and result files on GitHub[2], I can give something back to Nechakhin, D'Souza and Eger, and to their project.

# Data Availability Statement

The evaluation dataset created by Nechakhin and D'Souza is publicly accessible at https://data.uni-hannover.de/dataset/orkg-properties-and-llm-generated-research-dimensions-evaluation-dataset. The program code and prompts used in this study are available on the project's GitHub page: https://github.com/SandraSchaftner/orkg_property_extraction_using_gpt-3.5

# Conflicts of Interest

The author declares no conflicts of interest.

# Appendix

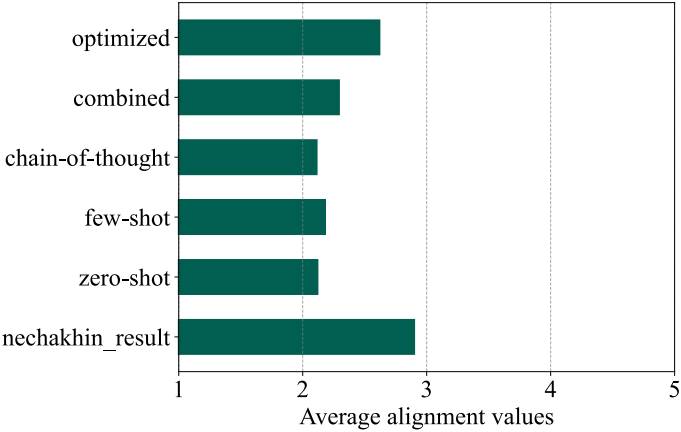

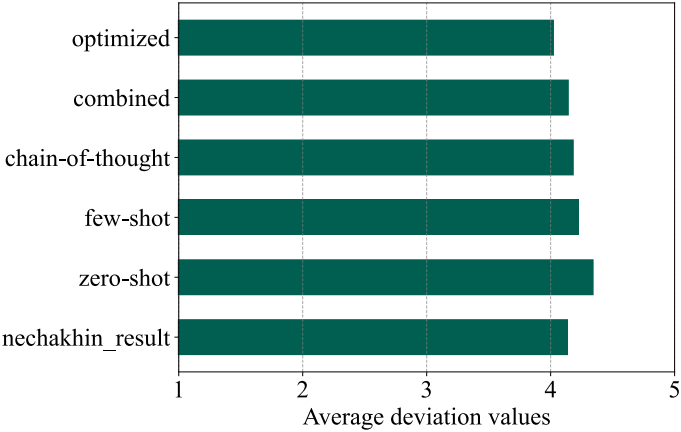

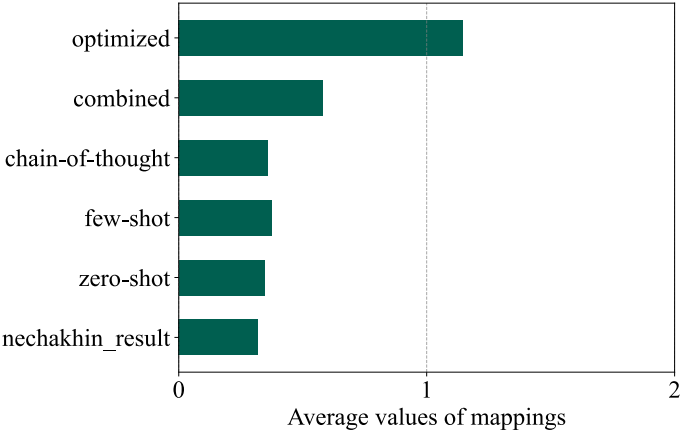

Figure A1. Bar charts of the evaluation scores for alignment, deviation and number of mappings. Regarding the alignment bar chart, it should be noted that the results of the preliminary study, referred to as *nechakhin_result*, could not be reproduced even approximately when using the same prompts (see the discussion in the first paragraphs of Section 5.1).

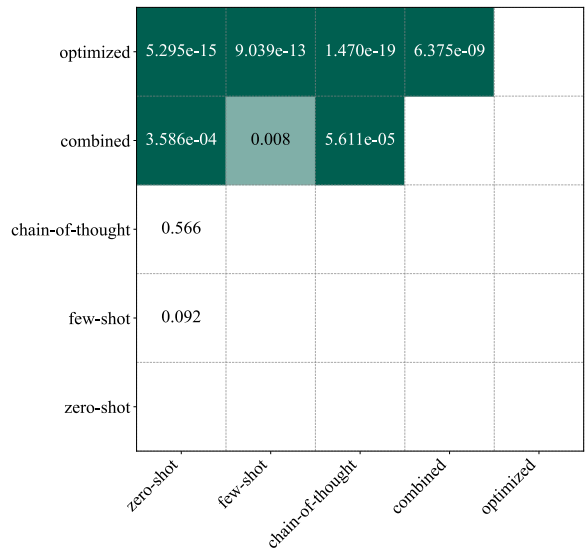

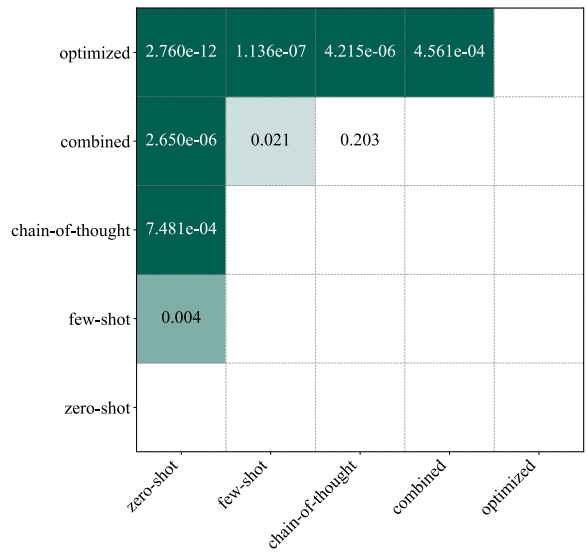

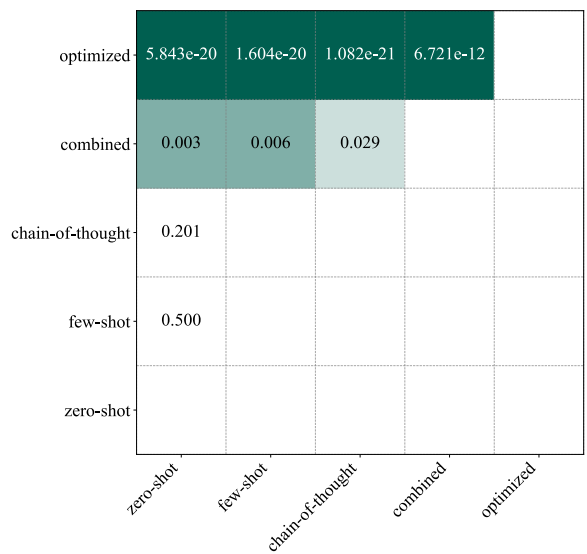

Figure A2. Heatmaps displaying the p-values from the sign tests for the results on alignment, deviation, and number of mappings. The heatmaps are to be interpreted as follows: *[The strategy listed on the vertical axis on the left]* is better by the p-value *[p-value]* than *[the strategy listed on the horizontal axis below]*. Only the cells that were calculated are filled in.